\documentclass[hyphens]{svproc}

\usepackage{url}

\usepackage{longtable}
\usepackage{booktabs}   
\usepackage{subcaption} 
\usepackage[unicode=true,colorlinks=false,linkcolor=black,hidelinks]{hyperref}
\usepackage{graphicx}

\usepackage{tikz-cd}
\usepackage{stackengine}
\usepackage{url}
\usepackage{url}

\IfFileExists{footnote.sty}{\usepackage{footnote}\makesavenoteenv{longtable}}{}
\usepackage{listings}

\usepackage{fancyvrb}

\DefineVerbatimEnvironment{Highlighting}{Verbatim}{commandchars=\\\{\}}

\newlength{\cslhangindent}
\newenvironment{cslreferences}%
  {\setlength{\parindent}{0pt}%
  \everypar{\setlength{\hangindent}{\cslhangindent}}\ignorespaces}%
  {\par}

\begin{document}
\mainmatter

\title{Multicloud API binding generation
from documentation}
\titlerunning{Multicloud binding
generation}

\author{ Michał J.
Gajda \inst{1} \and  Vitor Vitali
Barrozzi \inst{1} \and  Gabriel
Araujo \inst{1}}

\institute{Migamake Pte Ltd\\
\email{migamake@migamake.com}\\
\texttt{}\\ }

\maketitle

\hypertarget{introduction}{%
\section{Introduction}\label{introduction}}

We propose a solution for low-level
integration cloud service integration
among different providers without
additional support from the cloud API
vendor.

Multicloud service orchestration allows
leveraging the full power of
specialized, scalable best-of-a-kind
services in the cloud. Its advent
coincides with the so-called
\emph{no-code} and \emph{low-code}
solutions that enable rapid prototyping
of cloud applications by small teams of
startup companies. While excellent in
theory, but in practice SDKs usually
address only a few programming
languages, descriptions are untyped and
incomplete, API description languages
like Swagger/OpenAPI{[}20{]} are rarely
used. That means that initial
implementation is often an incoherent
mix of incompatible scripts, deployment
is only partly automated, and the source
code uses a variety of platforms with
varying level of automation.

An attempt to integrate these stumbles
upon the barrier of cloud APIs
documented in many different ways,
usually \emph{ad-hoc} and without
particular rigour. The API interface
bindings often contain implicit
assumptions, untyped JSON or text-based
bindings.

We solve this significant and essential
interoperability problem by
automatically parsing API documentation,
and then generating API bindings in any
chosen programming language\footnote{We
  currently generate Haskell{[}12{]}
  code, but developing generation to any
  other typed language is offered as a
  paid service upon request.}. The
additional efforts to support cloud API
bindings by code generation are limited
to single API and single language{[}1,
5, 11{]}, or require a significant
effort of handwriting Swagger/OpenAPI
declarations for entire API{[}30{]}, if
the vendor did not generate them.

\hypertarget{implementation}{%
\section{Implementation}\label{implementation}}

\hypertarget{pipeline}{}
\begin{figure*}[t]


\begin{tikzcd}[column sep=small]
\Centerstack{\textit{Data gathering}}                                          & [+15pt] & [-25pt] \textit{Parsing}                            & [-15pt]                                                               & \mathit{Code\ generation} &                                           \\
\rm{URLs} \arrow[d] \arrow[d, shift left=3] \arrow[d, shift right=3]           & \Centerstack{JSON parser} \arrow[rr] &         & \Centerstack{JSON Autotype type inference} \arrow[r]                          & \Centerstack{Argument and result type declarations} \arrow[dd] \arrow[rdd] & \Centerstack{Target library package} \\
\mathrm{Crawlers} \arrow[d] \arrow[d, shift right=3] \arrow[d, shift left=3]   & \Centerstack{Parameter descriptions} \arrow[rr] & & \Centerstack{Parameter types} \arrow[dd] \arrow[ru] \arrow[rd] &                           &                                           \\
\mathrm{Extractors} \arrow[d] \arrow[d, shift right=3] \arrow[d, shift left=3] & \Centerstack{HTTP request parser} \arrow[rrd] & &                                                                & \Centerstack{HTTP call function} \arrow[r] & \Centerstack{Typed source module} \arrow[uu] \\
\mathrm{CSV} \arrow[r] \arrow[ru, to path={-- ([xshift=3.3em]\tikztostart.east)
                                                       |- (\tikztotarget)
                                                      }, rounded corners]
                       \arrow[ruu, to path={-- ([xshift=2.8em]\tikztostart.east)
                                                       |- (\tikztotarget)
                                                      }, rounded corners]
                       \arrow[ruuu, to path={-- ([xshift=2.3em]\tikztostart.east)
                                            |- (\tikztotarget)
                                            }, rounded corners]                & \Centerstack{Use example parser} \arrow[rr, ""{coordinate, name=Z}] &  & \Centerstack{HTTP request type}                                &                           &
\end{tikzcd}
\caption{Data flow within the API binding code generation pipeline.}
\end{figure*}
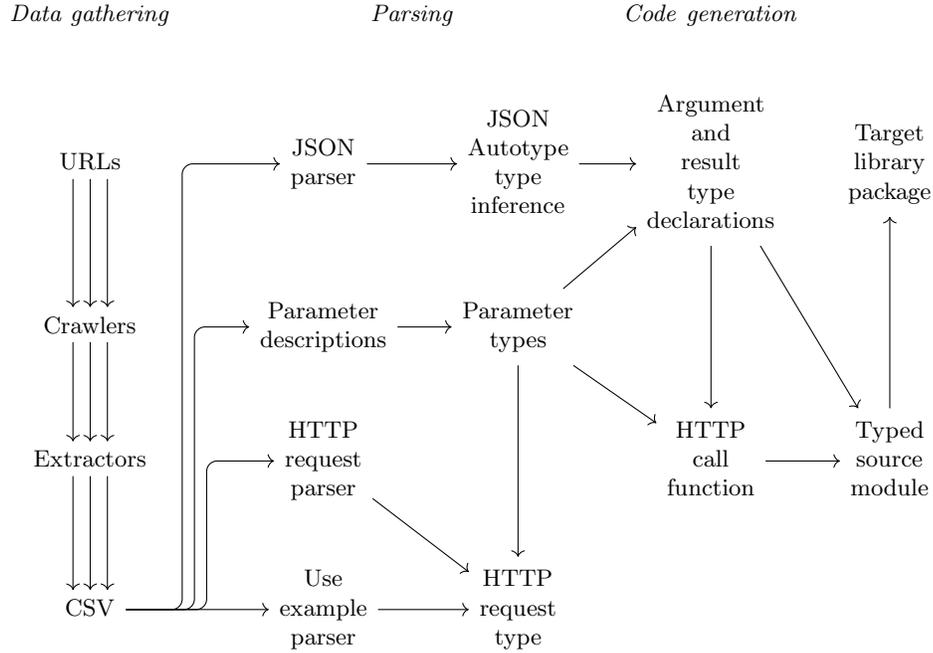

Our solution (\emph{HurlAPI}), is a
\emph{data pipeline}, divided into three
different stages: (I)~data gathering
from the web pages (II)~parsing and
analysis (III)~code generation.

\hypertarget{data-gathering}{%
\subsubsection{Data
gathering}\label{data-gathering}}

At the initial stage of \emph{data
gathering}, we gain a complete
description of each cloud API call with
Scrapy library{[}24{]} using Python. It
allows us to download HTML pages with
Chrome, then we examine page structure
with XPath{[}6{]}, CSS{[}25{]}, or
JQ{[}14{]} selectors to extract data in
a systematic manner. All this
information is written into tabular CSV
files{[}26{]}, with some columns
containing JSON objects, and carefully
validated.

\hypertarget{data-analysis}{%
\subsubsection{Data
analysis}\label{data-analysis}}

As part of a \emph{data analysis} stage
implemented in Haskell{[}12{]}, we parse
many possible formats: (a)~HTTP request
path description with variables (b)~cURL
command options (3) extracts from tables
of parameter descriptions. The parameter
descriptions are tagged with the
\emph{parameter passing
convention}\footnote{Cloud API call
  parameter passing conventions differ
  from binary function call conventions
  in this respect, and many different
  argument passing convention may be
  assigned for different parameters of
  the same call.}: (I)~as part of an
HTTP request path (II)~URL-encoded query
parameter (3)~part of a request body as
JSON or plain text{[}10{]} (4)~HTTP
header or~a cookie. The content of
parsed entries is carefully validated
and cross-checked for possible
inconsistencies. Every entry has a
separate list of errors, that are
reported per-record. While we only allow
100\% correct records to be used for
code generation, the failed records are
reported in detail. Summary statistics
of erroneous records reports on a
\emph{validation dashboard} that
indicates percentage correctness of
current data and allows us to assess the
overall health of data pipeline{[}9{]}.

\hypertarget{agile-data-pipeline-principles}{%
\paragraph{Agile data pipeline
principles}\label{agile-data-pipeline-principles}}

Data analytics pipeline goes beyond
previously described best practices in
agile data science{[}16{]} and also
draws from BCBS 239 best practices for
risk management in the financial
industry{[}4{]}.

The principles of our data pipeline
development process are (1) \emph{judge
by a final impact} -- prioritize
development of an entire pipeline to
judge issues by their impact on a final
product using data processing dashboard;
it allows us to focus our efforts on few
issues that have a significant impact on
final data\footnote{Only in the absence
  of more precise goals, the primary
  measure of impact is a frequency of an
  issue.} (2) \emph{record never
disappears} -- trace flow of records
over the entire data pipeline with
unique record identifers\footnote{In
  case records are merged, we also merge
  the identifiers.}; when filtering out
records, put them to alternate output,
so you can examine impact of each filter
(3) \emph{error is a tag or an alternate
output} -- assign error as one of many
\emph{tags} of the record, and then
filter by sorting error records to an
alternative output that requires similar
examination as final product; multiple
errors and warnings are attached to each
record that elucidate co-occurrence of
data quality and handling issues (4)
\emph{late filtering} -- delay filtering
when you have multiple data quality
criteria that can be run in parallel on
a single record; this makes it easy to
examine issue co-occurrence that is
common for faulty data (5)
\emph{universal data formats} -- data at
any stage of the pipeline is available
for examination as CSV files (6)
\emph{gradual record enrichment} with
additional information, so we can
examine all data related to the record
in a single row; preserve of existing
information, so inputs and outputs can
be quickly examined at any stage of
processing (7) an \emph{iteration
throughput} is considered as important
as an iteration speed, since number of
successfully processed records increases
a number of issues discovered during
each iteration, and we try to make the
processing of different records and
categories as independently as possible
(8) \emph{tagging potential gaps} with
errors or warnings, instead of assuming
total correctness of the input data and
the processing pipeline; this
facilitates data-based assessments of
completeness of the analysis (9) use
excerpts from \emph{real data as unit
tests} whenever possible to avoid
testing for issues rarely or never occur
in practice.\footnote{It also allows for
  principled treatment of the potential
  problems for which we do not have any
  practical test examples. We firmly
  anchor our analysis in naturally
  occurring data.}

The principles (1), (7-9) are all guided
by \emph{Zeno's principle}{[}13{]} of
extensive data processing, where sorting
data quality and processing issues by
the final impact on the final product
will show that most of the issues occur
in a relatively small number of records.
Fixing the first issues gives big
improvements, but getting to 100\%
accuracy needs much more work. We can
easily observe that moving from 80\% to
90\% of correctly processed records
takes about the same time as moving from
98\% to 99\%. Still, the gain in the
former case is more
substantial.\footnote{Which is kind of
  inverse Pareto{[}23{]} principle in
  data analytics. We can call it
  \emph{Zeno's principle} after the
  behaviour of the turtle in the Zeno's
  paradox{[}13{]}, since we move in
  smaller steps the closer we are to the
  complete correctness.} The principles
(3-5) aim to increase iteration
throughput in terms of simultaneously
detected data quality and processing
issues.

\hypertarget{code-generation}{%
\subsubsection{Code
generation}\label{code-generation}}

As the final stage, we \emph{generate
code} in a typed programming language.
We~start with a reference data structure
that lists necessary functions and type
declarations. This part is
language-independent except for function
generating the language-dependent
declaration identifiers themselves. The
binding generation proceeds with
templates that use these identifiers to
generate full code modules, and then
entire API binding package along with
its metadata; we use techniques
described in {[}2, 8, 15, 18{]}.
Following the best current practice, we
also attach links to the original
documentation website, which allows user
to cross-reference the information with
the original API documentation.

\hypertarget{conclusion}{%
\section{Conclusion}\label{conclusion}}

\hypertarget{limitations}{%
\subsubsection{Limitations}\label{limitations}}

When proposing a code generation
solution instead of handwritten code, it
is important to consider limitations
compared to manual processes. For a few
APIs, we need to implement specialized
components like AWS S3 chunked
signatures{[}27{]}, custom
authentication rolled out for
TransferWise~API{[}29{]}.\footnote{We
  agree with TransferWise{[}29{]}, that
  this is more secure than using plain
  secret like an access token, but each
  new authentication method needs
  special support code.} There are also
few (less than 2\% in entire
MS~Graph~API{[}19{]}) of API calls that
use custom argument passing. For
example, MS Graph uses custom DSL for
filtering by customizable \emph{extended
properties}{[}22{]}. Another example is
non-standard retry behaviour of
Backblaze~API{[}3{]}, that requires to
replace the access token upon receiving
503 response (service unavailable).
There are also sometimes bugs in the
documentation, which will cause the
generation of incorrect code. Some
companies provide only language-specific
SDKs{[}28{]}, instead of publicly
documenting their REST interfaces.
Luckily the situation improves with
bigger companies even providing live
debugging or live sandbox functionality
for the REST interface{[}7, 17, 21{]}.

\hypertarget{summary}{%
\subsubsection{Summary}\label{summary}}

We implemented the retargetable code
generator for cloud API bindings that
presents the following benefits:
(1)~provide a binding for thousands of
API calls within months; (2)~language
retargeting with little effort;
(3)~the~systematic approach allows easy
scaling to a number APIs; (4)~removes a
dependency on the cloud API provider
support; (5)~it~significantly reduces
maintained code base as compared with
handwritten cloud API bindings. We offer
to generate cloud API bindings for other
programming languages and other cloud
APIs as a paid service.

\hypertarget{bibliography}{%
\section*{Bibliography}\label{bibliography}}
\addcontentsline{toc}{section}{Bibliography}

\hypertarget{refs}{}
\begin{cslreferences}
\leavevmode\hypertarget{ref-amazonka}{}%
{[}1{]} Amazonka -- A comprehensive
Amazon Web Services SDK for Haskell:
2013.
\emph{\url{https://github.com/brendanhay/amazonka}}.

\leavevmode\hypertarget{ref-code-gen-prezi}{}%
{[}2{]} Art of industrial code
generation:
\emph{\url{https://www.migamake.com/presi/art-of-industrial-code-generation-mar-6-2019-uog-singapore.pdf}}.

\leavevmode\hypertarget{ref-backblaze-upload-error}{}%
{[}3{]} B2 Integration Checklist,
Uploading files:
\emph{\url{https://www.backblaze.com/b2/docs/integration_checklist.html}}.

\leavevmode\hypertarget{ref-bcbs-239}{}%
{[}4{]} Basel Committee on Banking
Supervision 2013. BCBS 239: Principles
for effective risk aggregation and risk
reporting.

\leavevmode\hypertarget{ref-boto}{}%
{[}5{]} Boto 3 - The AWS SDK for Python,
release 1.12.9: 2020.
\emph{\url{https://github.com/boto/boto3}}.

\leavevmode\hypertarget{ref-XPath}{}%
{[}6{]} Clark, J. and (eds.), S.D. 1999.
\emph{XML path language (XPath) version
1.0}. W3C.

\leavevmode\hypertarget{ref-transferwise-sandbox}{}%
{[}7{]} Full API Reference, Simulation:
\emph{\url{https://api-docs.transferwise.com/\#simulation}}.

\leavevmode\hypertarget{ref-code-gen}{}%
{[}8{]} Gajda, M.J. Do not give us a bad
name. Work in progress.

\leavevmode\hypertarget{ref-migamake-data-pipelines}{}%
{[}9{]} Gajda, M.J. Guidelines for agile
data pipelines. Work in progress;
Migamake Pte Ltd.

\leavevmode\hypertarget{ref-towards-better-union}{}%
{[}10{]} Gajda, M.J. Towards a more
perfect union type.
10.5281/zenodo.3929473.

\leavevmode\hypertarget{ref-gogol}{}%
{[}11{]} Gogol -- A comprehensive Google
Services SDK for Haskell: 2015.
\emph{\url{https://github.com/brendanhay/gogol}}.

\leavevmode\hypertarget{ref-haskell}{}%
{[}12{]} Haskell 2010 Language Report:
\emph{\url{https://www.haskell.org/definition/haskell2010.pdf}}.

\leavevmode\hypertarget{ref-zeno}{}%
{[}13{]} Huggett, N. 2019. Zeno's
Paradoxes. \emph{The Stanford
Encyclopedia of Philosophy}. Edward N.
Zalta, ed.
\url{https://plato.stanford.edu/archives/win2019/entries/paradox-zeno/};
Metaphysics Research Lab, Stanford
University.

\leavevmode\hypertarget{ref-jq}{}%
{[}14{]} jq is a lightweight and
flexible command-line JSON processor:
2012.
\emph{\url{https://stedolan.github.io/jq/manual/}}.

\leavevmode\hypertarget{ref-json-autotype-prezi}{}%
{[}15{]} JSON Autotype: Presentation for
Haskell.SG: 2015.
\emph{\url{https://engineers.sg/video/json-autotype-1-0-haskell-sg--429}}.

\leavevmode\hypertarget{ref-agile-data-science}{}%
{[}16{]} Jurney, R. 2017. \emph{Agile
data science 2.0: Building full-stack
data analytics applications with spark}.
O'Reilly Media, Inc.

\leavevmode\hypertarget{ref-mailgun-sandbox}{}%
{[}17{]} Mailgun Sandbox Domain
Explained: 2020.
\emph{\url{https://blog.mailtrap.io/mailgun-sandbox-tutorial/}}.

\leavevmode\hypertarget{ref-xml-typelift}{}%
{[}18{]} Michal J. Gajda, D.K. 2020.
Fast XML/HTML tools for Haskell: XML
Typelift and improved Xeno. Manuscript
in review, 10.5281/zenodo.3929548.

\leavevmode\hypertarget{ref-msgraph}{}%
{[}19{]} Microsoft Graph REST API v1.0
reference:
\emph{\url{https://docs.microsoft.com/en-us/graph/api/resources/domain?view=graph-rest-1.0}}.

\leavevmode\hypertarget{ref-openapi}{}%
{[}20{]} OpenAPI 3.0.2 Specification:
2018.
\emph{\url{https://swagger.io/docs/specification/about/}}.

\leavevmode\hypertarget{ref-alibaba-api-explorer}{}%
{[}21{]} OpenAPI Explorer:
\emph{\url{https://api.alibabacloud.com/}}.

\leavevmode\hypertarget{ref-azure-custom-dsl}{}%
{[}22{]} Outlook extended properties
overview: 2020.
\emph{\url{https://docs.microsoft.com/en-us/graph/api/resources/extended-properties-overview?view=graph-rest-1.0}}.

\leavevmode\hypertarget{ref-pareto}{}%
{[}23{]} Pareto, V. 1896. \emph{Cours
d'Economie politique}. Droz.

\leavevmode\hypertarget{ref-scrapy}{}%
{[}24{]} Scrapy:
\emph{\url{https://scrapy.org/}}.

\leavevmode\hypertarget{ref-CSSSelectors}{}%
{[}25{]} Selectors Level 3: 2011.
\emph{\url{https://www.w3.org/TR/2011/REC-css3-selectors-20110929/}}.

\leavevmode\hypertarget{ref-csv}{}%
{[}26{]} Shafranovich, Y. 2005. Common
Format and MIME Type for Comma-Separated
Values (CSV) Files. RFC 4180; RFC
Editor.

\leavevmode\hypertarget{ref-AWS-chunked-signature}{}%
{[}27{]} Signature Calculations for the
Authorization Header: Transferring
Payload in Multiple Chunks (Chunked
Upload) (AWS Signature Version 4): 2016.
\emph{\url{https://docs.aws.amazon.com/AmazonS3/latest/API/sigv4-streaming.html\#sigv4-chunked-body-definition}}.

\leavevmode\hypertarget{ref-smartfox-sdk}{}%
{[}28{]} SmartFoxServer 2X documentation
central: 2020.
\emph{\url{http://docs2x.smartfoxserver.com/}}.

\leavevmode\hypertarget{ref-transferwise-auth}{}%
{[}29{]} Strong customer authentication:
2020.
\emph{\url{https://api-docs.transferwise.com/\#payouts-guide-api-access}}.

\leavevmode\hypertarget{ref-swaggerhub}{}%
{[}30{]} SwaggerHub:
\emph{\url{https://swagger.io/tools/swaggerhub/}}.
\end{cslreferences}

\bibliography{multicloud-binding.bib}

\end{document}